\documentclass[pre,twocolumn,showpacs,amsmath,amssymb,preprintnumbers]{revtex4}

\usepackage{bm}

\newcommand\mA{{\bm{\mathsf A}}}
\newcommand\mC{{\bm{\mathsf C}}}
\newcommand\mD{{\bm{\mathsf D}}}
\newcommand\mF{{\bm{\mathsf F}}}
\newcommand\mG{{\bm{\mathsf G}}}
\newcommand\mL{{\bm{\mathsf \Lambda}}}
\newcommand\mI{{\bm{\mathsf I}}}
\newcommand\mJ{{\bm{\mathsf J}}}
\newcommand\mQ{{\bm{\mathsf Q}}}
\newcommand\mR{{\bm{\mathsf R}}}
\newcommand\mS{{\bm{\mathsf S}}}
\newcommand\mW{{\bm{\mathsf W}}}
\newcommand\mOmega{{\bm{\mathsf \Omega}}}

\newcommand\vj{{\bm j}}
\newcommand\vx{{\bm x}}
\newcommand\vy{{\bm y}}
\newcommand\vw{{\bm w}}
\newcommand\vv{{\bm v}}
\newcommand\vz{{\bm z}}
\newcommand\vxi{{\bm \xi}}

\newcommand\e{{\epsilon}}

\newcommand\Trace{\mathop{\rm Trace}\nolimits}
\newcommand\Det{\mathop{\rm Det}\nolimits}

\begin{document}

\preprint{to be published in Physical Review E}

\title{Fluctuation Properties of Steady-State Langevin Systems}
\date{August 24, 2007; Revised November 8, 2007}
\author{Jeffrey B. Weiss}
\affiliation{Department of Atmospheric and Oceanic Sciences,
  University of Colorado, Boulder, CO 80309}
\email{jeffrey.weiss@colorado.edu}

\begin{abstract}
Motivated by stochastic models of climate phenomena, the steady-state
of a linear stochastic model with additive Gaussian white noise is
studied. Fluctuation theorems for nonequilibrium steady-states provide
a constraint on the character of these fluctuations. The
properties of the fluctuations which are unconstrained by the
fluctuation theorem are investigated and related to the model
parameters. The irreversibility 
of trajectory segments, which satisfies a fluctuation theorem,
is used as a measure of nonequilibrium fluctuations. The moments of
the irreversibility probability density function (pdf) are found
and the pdf is seen to be non-Gaussian. The average irreversibility
goes to zero for short and long trajectory segments and has a
maximum for some finite segment length, which defines a
characteristic timescale of the fluctuations. The initial average
irreversibility growth rate is equal to the average entropy production
and is related to noise-amplification. 
For systems with a separation of deterministic timescales, modes with
timescales much shorter than the trajectory timespan and whose noise
amplitudes are not asymptotically large, do not, to first 
order, contribute to the irreversibility statistics, providing a
potential basis for dimensional reduction. 
\end{abstract}


\maketitle

\section{\label{sec:intro}Introduction}

Recent advances in nonequilibrium statistical mechanics have investigated
fluctuation theorems in a variety of
contexts \cite{EvansSearles2002}. The fluctuation theorem quantifies
the probability of finding fluctuations in nonequilibrium systems that
violate the Second Law of Thermodynamics. Fluctuation theorems take
many forms. The formulation we will focus on is in terms of the
probability of observing finite time trajectory segments of a system
\cite{SearlesEvans1999,Chernyak2006}. In this context, the fluctuation
theorem provides a constraint that such trajectory segments must
satisfy. Here we investigate the fluctuations in a nonequilibrium
steady-state governed by Langevin dynamics. We go beyond the
fluctuation theorem and study those properties of the fluctuations
that are not constrained by the fluctuation theorem.  These properties
are not generic. They depend on the details 
of the specific dynamical system, and we investigate the
relationship between the nonequilibrium fluctuations and the 
parameters defining the Langevin dynamics.

Our motivation for studying specific details of nonequilibrium
fluctuations comes from work in theoretical climate
dynamics. In recent years, linear stochastic dynamical systems have
been successfully used to model many phenomena in the climate system
such as
El-Ni\~no \cite{Penland93, PenlandSardeshmukh95, MooreKleeman96},
the North Atlantic Gulf Stream \cite{MooreFarrell93}, and a variety
of atmospheric phenomena
\cite{FarrellIoannou93, FarrellIoannou94, Newmanetal97,
  WhitakerSardeshmukh98, Weickmannetal00}. We shall refer to  
these phenomena as climate subsystems in that they are often
considered to be dynamical systems that are separable from the larger
climate system, at least on some set of spatial and temporal
scales. Since these fluctuations have macroscopic timescales, it is
important to investigate the character of individual fluctuations and
the statistics of their properties. 

In the work on climate subsystems, the focus has been on two
considerations: the utility  
of linear stochastic systems in forecasting \cite{Penland93, MooreKleeman96,
  PenlandMatrosova01}, and the potential 
for the deterministic part of the dynamics to amplify the random noise
\cite{FarrellgentheoryI, FarrellgentheoryII, Ioannou95,
 MooreKleeman96, Weiss2003}. It is often assumed that the large
amplitudes of these 
phenomena requires them to be the result of dynamical
instabilities. The recognition that deterministic dynamics can amplify
small noise forcing, which in meteorology goes back to Lorenz
\cite{Lorenz65}, provides an 
alternative view of such 
phenomena. This amplification occurs when the deterministic matrix is
non-normal \cite{FarrellgentheoryI, FarrellgentheoryII}. One common
critique of the noise-amplification view 
is that non-normality and the resulting amplification depends on the
subjective choice of coordinate system, and can be removed by an
appropriate coordinate transformation. Recently this objection has
been answered by noting that underlying the property of non-normality
is the more fundamental, coordinate invariant property of detailed
balance. Linear stochastic climate subsystem models share the property
that they violate detailed balance, and this is what is responsible
for the noise amplification \cite{Weiss2003}. Thus, a wide range of
phenomena in the climate system can be interpreted as fluctuations in
a nonequilibrium steady-state. For climate fluctuations such as
El-Ni\~no, understanding the character of the fluctuations is
extremely important. Further, due to global warming, the steady-state
is changing. Understanding how phenomena such as El-Ni\~no will change
as climate changes is a  major uncertainty in climate change
predictions \cite{vanOldenborgh2005}. Thus, improved understanding of
how nonequilibrium 
fluctuations depend on the properties of the steady-state could lead
to improvements in climate change forecasts.

As a concrete example, we will focus on El-Ni\~no. El-Ni\~no is a
coupled atmosphere-ocean phenomenon that is centered in the tropical
Pacific Ocean and has global impacts. One key aspect of El-Ni\~no is
that the atmosphere evolves on a faster timescale than the ocean. The 
turbulent dynamics of the atmosphere has a predictability limit of about
two weeks \cite{Vallis2006}. The ocean, on the other 
hand, has timescales of months. Thus, on the monthly timescale of
El-Ni\~no, the atmosphere is unpredictable and can be considered a
random forcing \cite{Penland96}. While this parameterization of fast
chaos as random noise is typically done empirically, there are
some theoretical results \cite{Leith1996,MTV1999,MTV2001,MTV2006}.

The phenomenon of El-Ni\~no is described in terms of the sea surface
temperature (SST) of the tropical Pacific. Although the SST is a
continuous field, both observations and models use a finite number $N$
of SST values. Thus, the state vector $\vx$ of the
system is an $N$-dimensional vector of real numbers representing the 
discretized SST field. Often, the dimensionality is reduced by truncating to
some number of leading modes. Typically, the mean SST is removed, so $\vx$
represents the SST anomaly and can be positive or negative. A linear
stochastic Langevin model for El-Ni\~no is then 
\begin{equation}
\label{eq:sde}
\frac{d \vx}{dt} = \mA \vx + \mF \vxi,
\end{equation}
where $\mA$ is an $N \times N$ real matrix representing the linear
deterministic dynamics, $\mF$ is an $N \times N$ real matrix
representing the noise forcing, $\vxi$ is $N$ dimensional Gaussian
white noise, $\langle \vxi(t)\vxi^T(s)\rangle = \mI \delta(t-s)$, where
superscript $T$ represents the transpose, $\mI$ is the
identity matrix, and the diffusion matrix is $\mD =
\mF \mF^T/2$. We require the system to have a steady-state, which 
implies that the deterministic dynamics is stable, i.e. all
eigenvalues of $\mA$ have a negative real part. From a dynamical
systems perspective, Eq.\ (\ref{eq:sde}) describes a stable linear
fixed point perturbed by additive Gaussian white noise. 
Eq.\ (\ref{eq:sde}) is the fundamental equation defining the
dynamics, and our goal is to describe the nonequilibrium fluctuations
in terms of $\mA$, $\mD$ and matrices derived from them. 

The Langevin dynamics, Eq.\ (\ref{eq:sde}), describes both
equilibrium and nonequilibrium steady states, depending on whether or
not detailed balance is satisfied. For most matrices $\mA$ and $\mD$,
detailed balance is violated and the system has a nonequilibrium
steady-state. This is the 
case for Langevin models of climate subsystems.
Detailed balance requires $\mA \mD - \mD \mA^T = 0$, and then
$\mOmega=\mA + \mD \mQ_0=0$, where 
$\mQ_0$ is the inverse of the steady-state covariance, defined
explicitly below, while both 
expressions are nonzero when detailed balance is violated. When
detailed balance is violated, the  steady-state distribution
$p_0(\vx)$ is maintained by a nonzero divergence-free probability
current $\vj(\vx) = \mOmega\vx p_0(\vx)$, where $\mOmega \vx$ is a
phase space velocity and $\mOmega$ can be interpreted as a matrix of
rotation frequencies. The probability current is divergence-free, and
thus $\Trace(\mOmega)=0$. The system satisfies detailed balance if and
only if there exists a 
coordinate system where $\mA$ and $\mD$ are both diagonal. Thus
systems described by Eq.\ (\ref{eq:sde}) in detailed balance can be
transformed into a collection of uncoupled one-dimensional systems,
while those violating detailed balance have an essential
multi-dimensional character. 

\section{\label{sec:ft}The fluctuation theorem}

The fluctuation theorem can be written in terms of the probability of
observing trajectory segments. Consider a long
trajectory $\vx$ and choose a time interval $t$. For any two
states $\vx_0$ and $\vx_1$, we define the trajectory probability
$p(\vx_0,\vx_1,t)$ as the probability of finding a trajectory
segment within the long trajectory that begins at $\vx_0$ and ends at
$\vx_1$ a time $t$ later. This definition differs from the trajectory
probability used in some previous studies \cite{Chernyak2006,
  Seifert2005} in that it depends only on the endpoints of the
trajectory segment and not on the full trajectory. By relying only on
the endpoints of the trajectory segment, this definition may be better
suited to discretely sampled data, such as found in climate records.

The time-reversed trajectory segment,
one starting at $\vx_1$ and ending at $\vx_0$, has a probability
$p(\vx_1,\vx_0,t)$. For simplicity we shall drop the dependence on the
time interval $t$ when it causes no confusion. The irreversibility
$r(\vx_0,\vx_1,t)$ of a trajectory segment with initial state $\vx_0$
and final state $\vx_1$ is defined by   
\begin{equation}
\label{eq:rdef}
r(\vx_0,\vx_1) = \ln\frac{p(\vx_0,\vx_1)}{p(\vx_1,\vx_0)}.
\end{equation}
Thus, $r\equiv 0$ says that one is equally
likely to find forward and reverse trajectory segments and the system
is reversible, while $r\ne 0$
says that one can distinguish forward from reverse
trajectories segments.  The statistics of $r$ over a long trajectory
corresponds to the intuitive notion of irreversibility: they quantify
how well one is able to distinguish a long trajectory from the same
trajectory going backwards in time.

The fluctuation theorem is now a direct result of the definition of
$r$ \cite{Chernyak2006}. Eq.\ (\ref{eq:rdef}) implies that $r(\vx_1,\vx_0) =
-r(\vx_0,\vx_1)$, and $p(\vx_0,\vx_1) = \exp(r(\vx_0,\vx_1))
p(\vx_1,\vx_0)$. 
Then the probability $p_r(r)$ of finding a 
trajectory segment with irreversibility $r$ is
\begin{eqnarray}
\label{eq:flucderiv}
p_r(r) &=& \int d^N \vx_0 \,d^N \vx_1 \,p(\vx_0,\vx_1)
\delta(r(\vx_0,\vx_1)-r)\nonumber\\
 &=&\int d^N \vx_0 \,d^N \vx_1 \,
e^{-r(\vx_1,\vx_0)} 
\nonumber\\ &&\qquad\times p(\vx_1,\vx_0)\delta(r(\vx_1,\vx_0)+r)\nonumber\\
&=& e^r p_r(-r).
\end{eqnarray}
The final line in Eq.\ (\ref{eq:flucderiv}) is the fluctuation
theorem relating the 
probability of finding positive and negative irreversibilities.
The fluctuation theorem thus puts a constraint on $p_r(r)$, but does
not completely determine its functional form.

\section{\label{sec:irrtrajsec}The irreversibility of trajectory segments}

For the linear stochastic Langevin dynamics, Eq.\ (\ref{eq:sde}),
the probability of finding a trajectory segment with a given
irreversibility $p_r(r)$ can be written in terms of the basic
parameters of the dynamics. We begin by defining two more traditional
probability distributions: the steady-state probability $p_0(\vx)$ of finding the
system in state $\vx$, and the transition probability $p(\vx_1,t |
\vx_0)$ of finding the system in state $\vx_1$ conditioned on the
system being in state $\vx_0$ a time $t$ earlier. The trajectory
probability is then
\begin{equation}
\label{eq:trajpdf}
p(\vx_0,\vx_1,t) = p(\vx_1,t | \vx_0)p_0(\vx_0).
\end{equation}
Because the system is linear with additive Gaussian white noise, these
probabilities are also Gaussian and can be explicitly written in terms
of the covariance of the dynamics. 

A stochastic trajectory of Eq.\ (\ref{eq:sde}) starting at $\vx(0)$
can be written as 
\begin{equation}
\label{eq:trajsoln}
\vx(t) = e^{\mA t}\vx(0) + \int_0^t ds \, e^{\mA(t-s)} \mF \vxi.
\end{equation}
Since we only consider additive noise, the above integral is the same
in both Ito and Stratonovich interpretations.
The time-dependent covariance matrix $\mC_t$ for 
$\Delta\vx = \vx_1-\exp(\mA t)\vx_0$, the difference
between the stochastic trajectory and its deterministic counterpart is
\begin{eqnarray}
\label{eq:Ct} 
\mC_t &=& \langle
\Delta\vx \Delta\vx^T\rangle
\nonumber\\ 
&=& 2\int_0^t e^{\mA(t-s)} \mD e^{\mA^T(t-s)},
\end{eqnarray}
and the steady-state covariance $\mC_0 = \lim_{t\to\infty} \mC_t$
satisfies the relation
\begin{equation}
\label{eq:fdr}
\mA \mC_0 + \mC_0 \mA^T + 2\mD = 0.
\end{equation}
Writing $\mC_0$ as
\begin{equation}
\label{eq:C0int}
\mC_0 = 2 \int_{-\infty}^0 ds \, e^{-\mA s} \mD e^{-\mA^T s},
\end{equation}
allows one to obtain the useful relation
\begin{equation}
\label{eq:Cidentity}
e^{\mA t}\mC_0 e^{\mA^T t}= \mC_0 -\mC_t.
\end{equation}
Define the inverse covariance matrices, sometimes called the
concentration, by $\mQ_t = \mC_t^{-1}$, $\mQ_0 = \mC_0^{-1}$.

The steady-state probability is
\begin{equation}
\label{eq:sspdf}
p_0(\vx) = {\cal N}_0\exp[-\vx^T \mQ_0\vx/2],
\end{equation}
with normalization ${\cal N}_0 = 1/{\sqrt{(2\pi)^N \Det(\mC_0)}}$. 
The transition probability is
\begin{equation}
\label{eq:transpdf}
p(\vx_1,t |\vx_0) = {\cal N}_{t}
\exp[-\Delta\vx^T \mQ_t \Delta\vx/2],
\end{equation}
where the normalization is ${\cal N}_t= 1/\sqrt{(2\pi)^N
\Det(\mC_t)}$. The transition  
probability is thus a Gaussian distribution with covariance $\mC_t$
distributed around the endpoint of the deterministic trajectory
segment beginning at $\vx_0$.

The trajectory probability depends on both the initial and final
states. We thus consider a $2N$-dimensional state-space
\begin{equation}
\label{eq:zdef}
\vz = 
\begin{pmatrix}
\vx_0 \\ \vx_1
\end{pmatrix}.
\end{equation}
The forward trajectory probability is Gaussian and is given by
\begin{equation}
\label{eq:ptraj}
p(\vx_0,\vx_1) = {\cal N} \exp[-\vz^T \mR_{01} \vz/2],
\end{equation}
with normalization ${\cal N} = 1/[(2\pi)^N \sqrt{\Det(\mC_0
\mC_t)}]$, and the concentration of the trajectory probability
$\mR_{01}$ is the $2N \times 2N$ matrix
\begin{equation}
\label{eq:R01def}
\mR_{01} = 
\begin{pmatrix}
e^{\mA^T t}\mQ_t e^{\mA t} + \mQ_0 & \quad -e^{\mA^T t}\mQ_t\\
-\mQ_t e^{\mA t} & \quad \mQ_t
\end{pmatrix}.
\end{equation}
The reverse trajectory probability is
\begin{equation}
\label{eq:prevtraj}
p(\vx_1,\vx_0) = {\cal N} \exp[-\vz^T \mR_{10} \vz/2],
\end{equation}
where the concentration of the reverse trajectory probability is
related to that of the forward trajectory probability by swapping
initial and final states:
\begin{equation}
\label{eq:R10def}
\mR_{10} = \mJ \mR_{01} \mJ = 
\begin{pmatrix}
\mQ_t & \quad -\mQ_t e^{\mA t} \\
-e^{\mA^T t}\mQ_t& \quad e^{\mA^T t}\mQ_t e^{\mA t} + \mQ_0 
\end{pmatrix},
\end{equation}
with
\begin{equation}
\label{eq:Jdef}
\mJ = 
\begin{pmatrix}
0 & \mI\\
\mI & 0 
\end{pmatrix},
\end{equation}
and $\mI$ is the $N \times N$ identity matrix.

The irreversibility $r$ is now simply obtained from Eq.\
(\ref{eq:rdef}), $r(\vz) = \vz^T \mR 
\vz/2$ with the irreversibility concentration matrix $\mR = \mR_{10} -
\mR_{01}$. 
One can show that when detailed balance is
satisfied, the irreversibility concentration matrix $\mR$ is
identically zero and the system is reversible. 
The irreversibility can be be described by the eigenvalues and eigenvectors
of $\mR$. A $2N$-dimensional eigenvector 
$\vv_n$ with eigenvalue $\lambda_n$ can
be written in terms of an $N$-dimensional initial endpoint $\vv_{n0}$
and $N$-dimensional final endpoint $\vv_{n1}$, $\vv_n^T =
(\vv_{n0}^T,\vv_{n1}^T)$. Due to the structure of $\mR$, each
eigenvector has a corresponding time-reversed eigenvector with
reversed endpoints, $(\vv_{n1}^T,\vv_{n0}^T)$, with eigenvalue
$-\lambda_n$.  

\section{\label{sec:irrpdf}The irreversibility distribution}

The probability density function of irreversibility $p_r(r)$ can be
obtained in terms of its characteristic function $\hat p_r(k)$ 
\begin{eqnarray}
\label{eq:chardef}
\hat p_r(k) &=& \langle e^{i k r}\rangle = \int dr \, e^{ikr}
p_r(r)\nonumber\\
p_r(r) &=& \frac{1}{2 \pi} \int dk\, e^{-ikr} \hat p_r(k).
\end{eqnarray}
Using the result from the previous section, the irreversibility pdf
can be written as
\begin{equation}
\label{eq:prdef}
p_r(r) = \int d^{2N}\vz \, \delta(r - \vz^T \mR \vz/2) p(\vx_0,\vx_1).
\end{equation}
Writing the delta-function in terms of its Fourier transform
$\delta(r) = 1/(2\pi) \int dk \, \exp[i k r]$, changing the order
of integration, and comparing with Eq.\ (\ref{eq:chardef}) gives 
the characteristic function
\begin{equation}
\label{eq:pchar1}
\hat p_r(k) = {\cal N}\int d^{2N}\vz \,
\exp[-\vz^T (\mR_{01}- i k \mR)\vz/2].
\end{equation}
Since $\mR$ is symmetric and $\mR_{01}$ is positive
definite, the above expression can be simplified using the dual conjunctive
diagonalization \cite{MatrixRefMan}. This diagonalizes the matrix in
the exponential, transforming the expression into a product of
one-dimensional integrals that can be easily carried out. Define the matrix
$\mW=\mR_{01}^{-1}\mR$, define the matrix $\mS^{-1}$ to be the matrix of
eigenvectors of $\mW$, and define the diagonal matrix $\mL$ to be the
matrix whose elements are the $2N$ eigenvalues of $\mW$,
$\lambda_n$. Then $\mR_{01} =\mS^T \mS$ and $\mR = \mS^T \mL \mS$ and
Eq.\ (\ref{eq:pchar1}) can be integrated to obtain the 
characteristic function of the irreversibility pdf
\begin{equation}
\label{eq:pr3}
\hat p_r(k) = \frac{1}{\sqrt{\Det(\mI_{2N}- i k \mL)}} =
\frac{1}{\prod_{n=1}^{2N} \sqrt{1 - i k \lambda_n}},
\end{equation}
where we have made use of the fact that
$\Det(\mR_{01})\Det(\mC_0\mC_t) = 1$. 
Thus, the characteristic function depends on the eigenvalues of
$\mW$. 

The covariance of the 
the forward trajectory probability $\mR_{01}^{-1}$, needed to compute
$\mW$, can be written in a relatively simple form using Eq.\
(\ref{eq:Cidentity})  
\begin{equation}
\label{eq:r01inverse}
\mR_{01}^{-1} = 
\begin{pmatrix}
\mC_0 & \mC_0 e^{\mA^T t}\\
e^{\mA t} \mC_0 & \mC_0
\end{pmatrix}.
\end{equation}

We have thus obtained a closed-form expression for $\hat p_r(k)$ in
terms of the eigenvalues of $\mW$, which is an explicit function
of the deterministic dynamics $\mA$, the steady-state and finite-time
covariance matrices $\mC_0$ and $\mC_t$, and their inverses. The
covariance matrices depend only on the deterministic dynamics and the
diffusion matrix. The full pdf of irreversibility is then
obtained by integrating Eq.\ (\ref{eq:chardef}) using Eq.\
(\ref{eq:pr3}). No closed form solution for the integral has been
found. 

The eigenvalues of $\mW$ have an interesting interpretation. In the
coordinate system $\hat\vz = \mS \vz$, the trajectory probability
$p(\hat\vz) \sim \exp{\left(-\left|\hat\vz\right|^2/2\right)}$, and the
irreversibility is $r = \sum_i \lambda_i \hat z_i^2/2$. Thus, the
eigenvalue $\lambda_i$ of $\mW$ gives the 
irreversibility weighting for coordinate $i$ in a coordinate system
where the trajectory probability is isotropic and has unit
covariance. 

The moments of the irreversibility pdf can be 
obtained by considering the cumulants of the pdf \cite{AbramSteg}. The
cumulant function $g(k) = \log \hat p(k)$ is 
\begin{equation}
\label{eq:cumfundef}
g(k) = -\frac{1}{2}\sum_{n=1}^{2N} \log(1 - i k
\lambda_n),
\end{equation}
and the cumulants are then
\begin{eqnarray}
\label{eq:cumderiv}
\kappa_m &=& (-i)^m \left .\frac{d^m g(k)}{dk^m}\right|_{k=0}\nonumber\\
&=& \frac{(m-1)!}{2}\Trace(\mW^m).
\end{eqnarray}
Using the relationship between cumulants and centered moments gives
the moments of the pdf. The first four centered moments are
\begin{eqnarray}
\label{eq:fourmom}
\langle r \rangle &=& \frac{1}{2} \Trace(\mW),\nonumber\\
\langle \left(r - \langle r \rangle\right)^2\rangle&=&
\frac{1}{2}\Trace(\mW^2),\nonumber\\
\langle \left(r - \langle r
\rangle\right)^3\rangle&=&\Trace(\mW^3),\nonumber\\ 
\langle \left(r - \langle r \rangle\right)^4\rangle&=&
3\Trace(\mW^4)+\frac{3}{4}\Trace^2(\mW^2).
\end{eqnarray}
The irreversibility pdf is not, in general, Gaussian. The
distribution is skewed and has kurtosis different from the 
Gaussian value. Such non-Gaussian distributions have been previously
seen in a variety of nonequilibrium systems.\cite{CarberryEtAl2004,
  Ritort2004, BlickleEtAl2006,CrooksJarzynski2007}.   
Considering the definition of $r$, Eq.\
(\ref{eq:rdef}), and writing moments of the irreversibility as   
\begin{equation}
\label{eq:avgrpos}
\langle r^n \rangle = \int d^N \vx_0 \,d^N \vx_1 \,r^n(\vx_0,\vx_1)
p(\vx_0,\vx_1), 
\end{equation}
one sees that all (non-centered) moments are non-negative. The average
irreversibility can be simplified to 
\begin{equation}
\label{eq:avgr}
\langle r \rangle = \Trace(\mC_0\mQ_t(\mI - e^{2\mA t}) - \mI).
\end{equation}

It is interesting to consider how the irreversibility scales with the
strength of the noise. Consider taking a system and increasing the
diffusion matrix by a constant factor $\alpha$. Then the covariance matrices
increase by $\alpha$, and the concentration matrices are scaled
by $1/\alpha$. The irreversibility for a segment with fixed endpoints
scales by $1/\alpha$, but since the covariance scales by $\alpha$, the
probability of finding those endpoints scales. As a result, the
irreversibility pdf is unchanged by the scaling. 
Note that this invariance is only valid for multiplication by a
scalar. A matrix transformation of the noise will, in general, change
the dynamics and the irreversibility. 
The irreversibility
is coordinate invariant, so we are free to consider any convenient
coordinate system. If one considers a coordinate system where $\mD$ is
diagonal, obtainable by an orthogonal transformation, then the system
can be considered to be coupled to $N$ thermal reservoirs with temperatures
given by the eigenvalues of $\mD$. Then the irreversibility statistics
are unchanged by changing the temperature of all heat baths by a
constant factor. Further, if one considers the coordinate system where
$\mD = \mI$, obtainable by a non-orthogonal transformation, then the
system can be considered to have all $N$ degrees of freedom coupled to
a single thermal reservoir with unit temperature. This coordinate
system makes explicit the fact that all coordinate-invariant
properties, including the irreversibility, are independent of the
temperature of this single reservoir. Thus irreversibility is neither
a measure of the temperature of the reservoir, nor a measure of the
amplitude of the fluctuations, but rather is related to the
amplification of the noise above that seen in equilibrium.

\section{\label{sec:timedep}Time dependence of average irreversibility}

The time dependence of the average irreversibility gives information
about the time dependence of the fluctuations. 
Because the dynamics is stable, the following properties hold: the
eigenvalues of $\mA$ all have 
negative real part, $\lim_{t\to\infty} e^{\mA t} = 0$,
$\lim_{t\to\infty} \mC_t = \mC_0$, and $\lim_{t\to\infty} \mQ_t =
\mQ_0$. Thus, Eqs. (\ref{eq:R01def}) and (\ref{eq:R10def}) give
\begin{equation}
\label{eq:limR01R10}
\lim_{t\to\infty} \mR_{01} = \lim_{t\to\infty} \mR_{10} =
\begin{pmatrix}
\mQ_0 & 0\\
0& \mQ_0 
\end{pmatrix}.
\end{equation}
This is easily understood in that as the time becomes large, the
initial and final states become uncorrelated and the trajectory
probability is merely the product of the steady-state probabilities of
the initial and final states. As a result, $\lim_{t\to\infty} \mR =
0$, and the irreversibility of long trajectory segments goes to zero.

For short times $t = \tau \ll 1$, one can use the asymptotic
expansion for $e^{\mA \tau}$ to obtain expressions for
$\mC_\tau$ and $\mQ_\tau$.  This allows one to 
write the average irreversibility as
\begin{equation}
\label{eq:avgrtau}
\langle r \rangle = \frac{\tau}{2}\Trace\left[\mA \mC_0
\left(\mA^T \mD^{-1} - \mD^{-1}\mA\right)\right] + O(\tau^2).
\end{equation}
Thus, as $t \to 0$, $\langle r \rangle \to 0$. Further, for short
times the dependence of $\langle r \rangle$  on the violation of
detailed balance, which is equivalent to $\mA^T \mD^{-1} - \mD^{-1}\mA
\ne 0$, is manifest.  

Thus, the average irreversibility is positive semi-definite for all
time and goes to zero as $t\to 0$ and $t\to \infty$. This means that
either $\langle r \rangle (t) = 0$ for all $t$ and the system is in
equilibrium, or there is a time $t^*$ where the average
irreversibility reaches a maximum. This defines a characteristic
timescale of the irreversible fluctuations. 

Fluctuation theorems are often formulated in terms of the entropy
production \cite{EvansSearles2002}. Chernyak, et al.,
\cite{Chernyak2006} speculate that for multi-dimensional systems such
as Eq.\ (\ref{eq:sde}), the time average entropy increase in the
thermal reservoirs coupled to the system over a time $t$ is, in our notation,
\begin{equation}
\label{eq:entropyprod}
\Sigma_t = \frac{1}{t}\int_0^t \, d\vx^T \mD^{-1} \mOmega\vx,
\end{equation}
where the integral is interpreted in the Stratonovich sense.
Using the equation of motion Eq.\ (\ref{eq:sde}) to write $d\vx$ in
terms of $dt$ and taking the average one obtains  
\begin{eqnarray}
\label{eq:entsoln}
\langle \Sigma_t\rangle &=& \frac{1}{t}\int_0^t ds \, \langle \vx^T(s) \mA^T
\mD^{-1}\mOmega \vx(s)\rangle \nonumber\\
&\quad& +  \frac{1}{t}\int_0^t ds \, \langle \vxi^T(s) \mF^T
\mD^{-1}\mOmega \vx(s)\rangle .
\end{eqnarray}
The first term is simplified by considering $\vx(s)$ to be in the
steady-state so $\langle  x_i(s)x_j(s)\rangle=C_{0_{ij}}$. The second
term can be simplified to $\Trace(\mOmega)=0$. Manipulating the result
using the definition of $\mOmega$ and Eq.\ (\ref{eq:fdr}) one finds that 
the average steady-state rate of entropy production is
identical to the zero-time growth rate of the average trajectory
irreversibility given by Eq.\ (\ref{eq:avgrtau}),  
\begin{equation}
\label{eq:entravgr}
\langle\Sigma_t\rangle = \left. \frac{d \langle
  r\rangle(t)}{dt}\right|_{t=0}.
\end{equation} 

The entropy production can be related to the noise amplification that
occurs when detailed balance is violated. First consider
one-dimensional dynamics governed by deterministic dynamics $A < 0$,
diffusion constant $D > 0$, and steady-state covariance $C_0 > 0$. Then Eq.\
(\ref{eq:fdr}) gives $-(A C_0/D + 1)=0$. The negative sign is chosen so
that amplification will correspond to a positive number. If, through
some additional 
forcing, the covariance were increased to $C'>C_0$, then one measure of
that additional amplitude is $-(A C'/D+1) > 0$. In multi-dimensional
systems, increased variance occurs not through some additional
forcing, but through violation of detailed balance
\cite{Weiss2003}. One analogous measure for noise amplification is the
nondimensional gain matrix $\mG  = -(\mA\mC_0\mD^{-1} + \mI)$. Using
Eq.\ (\ref{eq:fdr}) allows one to write $\Sigma_t$ as
\begin{equation}
\label{eq:entrnoiseamp}
\langle\Sigma_t\rangle = \Trace\left[\mA \mG\right].
\end{equation} 
The entropy production rate has units of 1/time, and the
timescale of the system is set by the deterministic dynamics $\mA$,
Thus, $\mA \mG$ is a matrix measure of the noise-amplification per
unit time, whose trace gives the steady-state rate of entropy production.

\section{\label{sec:slowreduc}Reduction to slow modes} 

Complex spatio-temporal systems are typically high-dimensional systems
with a wide range of time-scales. One often would like to reduce the
system down to a more 
manageable number of degrees of freedom. There are several common
truncations but none are entirely satisfactory. Here we explore
dimensional reduction based on a separation of timescales and examine the
effect on irreversibility.

We have already seen that for times longer than the longest
deterministic timescale the irreversibility becomes zero. It is
thus not unreasonable to hypothesize that those degrees-of-freedom
whose timescale is much shorter than the timespan of a trajectory segment
will have no affect on its irreversibility. If this were the case, then
in considering the irreversibility of trajectory segments of a
particular timespan, one could reduce the dimensionality to 
those degrees of freedom whose timescales are similar to or longer
than the segment timespan. We now show that when there is 
a separation of timescales in the dynamics this is indeed the case and
the fast modes have a small effect on the irreversibility statistics.

Climate subsystems often have slow modes which comprise a
reduced dimensional subspace. Here we shall investigate the
consequences of the deterministic dynamics having a separation of
timescales. Note that by modeling the system as an $N$-dimensional
stochastic dynamical system 
we have already assumed a separation into $N$ slow modes, considered
to be deterministic, and unresolved fast modes which are parameterized
as random noise. Now we decompose the slow deterministic modes into slower
modes and faster modes. The faster modes are fast compared to the slower
modes, but still slow compared to the random noise. More 
specifically, assume that the $N$ eigenvalues $\lambda_i'$ of $\mA$ can
be divided into two groups, $N_s$ slow modes and $N_f=N-N_s$ fast modes
where $Re(\lambda_i') \sim O(1)$ for $1\le i \le N_s$, and
$Re(\lambda_i') \sim O(1/\epsilon)$ for $N_s < i \le N$, and
$\epsilon \ll 1$. Further, we restrict ourselves to times $t$ that are
order one, so that $Re(\lambda_{fast}') t \sim O(1/\epsilon)$. We shall
rescale the fast eigenvalues so that the small parameter is explicit,
$\lambda_{slow}=\lambda_{slow}'$,
$\lambda_{fast}=\lambda_{fast}'\epsilon$, and the real parts are all $O(1)$.

We will work in a coordinate system where $\mA$ is diagonal:
\begin{equation}
\label{eq:Apartition}
\mA =
\begin{pmatrix}
\Lambda_s& 0\\
0& \Lambda_f/\epsilon
\end{pmatrix},
\end{equation}
where $\Lambda_s$ is a diagonal matrix of slow eigenvalues and
$\Lambda_f$ is a diagonal matrix of scaled fast eigenvalues. Since the
eigenvalues will typically be complex, the coordinate transformation
is also complex. The various equations presented above must then be
modified by changing transpose operators to adjoint operators,
indicated by $\dagger$. For more
details on coordinate transformations in stochastic linear systems see
\cite{Weiss2003}. In these coordinates, the state space variable $\vx$
can be decomposed into slow variables $\vy$ and fast variables
$\vw$: $\vx^\dagger = ( \vy^\dagger, \vw^\dagger )$. 

In these diagonal-$\mA$ coordinates, the covariance matrix can be obtained
by direct integration of Eq.\ (\ref{eq:Ct}) to obtain 
\begin{equation}
\label{eq:Csolnfs}
C_{t_{ij}} = \frac{2 D_{ij}}{\lambda_i' + \lambda_j'^*}
\left[\exp\left(\left(\lambda_i' +
\lambda_j'^*\right)t\right)-1\right], 
\end{equation}
where star denotes complex conjugate and the eigenvalues are
unscaled. For order one times, $\exp(\lambda_{fast}' t) \sim
\exp(-1/\epsilon)\approx 0$ and the deterministic Green function is
\begin{equation}
\label{eq:greenfs}
e^{\mA t} = 
\begin{pmatrix}
e^{\Lambda_s t}&0\\
0&0
\end{pmatrix}.
\end{equation}
Then the covariance matrices take the form 
\begin{equation}
\label{eq:C0tfs}
\mC_t=
\begin{pmatrix}
\mC_{tyy}& \e\mC_{0yw}\\
\e\mC_{0wy}& \e\mC_{0ww}\\
\end{pmatrix},
\quad
\mC_0=
\begin{pmatrix}
\mC_{0yy}& \e\mC_{0yw}\\
\e\mC_{0wy}& \e\mC_{0ww}\\
\end{pmatrix},
\end{equation}
where the subscripts y/w indicate fast/slow modes, the subscripts
t/0 indicate whether the covariance sub-matrix is time-dependent or
steady-state, and all submatrices are $O(1)$.  Thus, the
time-dependent covariance matrix is decomposed into an $O(1)$
slow-slow covariance which depends on time, and its complement which is
small and equal to the steady-state value.
Note that this analysis assumes that all components of the diffusion
matrix $\mD$ are $O(1)$. It may be the case that the fast modes have
larger noise than the slow modes. If the noise in the fast modes is
$O(1/\epsilon)$ then this asymptotic expansion breaks down. 

Using this decomposition in the equations leading to $\mR$ and $\mW$,
and making frequent use of matrix identities for block matrices
leads to the following results. For any given trajectory segment with
endpoints $\vx_0^\dagger=(\vy_0^\dagger,\vw_0^\dagger)$,
$\vx_1^\dagger=(\vy_1^\dagger,\vw_1^\dagger)$, the irreversibility
$\vz^\dagger\mR\vz$ depends on both the fast and slow
variables. On the other hand, the characteristic function for the
irreversibility pdf depends to first order only on the slow
variables. Thus the statistics of the irreversibility are, to lowest
order, unaffected by truncating the fast variables. These two results
may seem contradictory. However, the fast eigenvalues have two effects: the
correlations decay rapidly and, as seen in Eq.\ (\ref{eq:C0tfs}),
the covariance of the fast modes is small. Thus, the typical size
of the fast variables is small and its contribution to the
irreversibility is a higher-order effect. While large rare
fluctuations in the fast modes do effect the irreversibility of
isolated trajectory segments, they do not, to lowest order, impact the
statistics. 
Thus, in considering irreversibility, and provided the diffusion
matrix is $O(1)$, one can safely
neglect fast modes and reduce the dimensionality of the dynamics to
just the slow modes of the system.

\section{\label{sec:Discussion} Discussion} 

Fluctuation theorems in nonequilibrium systems have focused attention
on the irreversibility of trajectory segments, and provide a
constraint for their distributions. Theories of climate subsystems
provide motivation for analyzing the nonequilibrium fluctuations of
linear stochastic dynamical systems in more detail. Here we combined
these two perspectives and obtained information about the trajectory
irreversibility that is not constrained by the fluctuation theorem.

For linear stochastic dynamics with additive Gaussian white noise, the
irreversibility was shown to be governed by an irreversibility
concentration matrix $\mR$ which is expressed in terms of the
fundamental matrices governing the dynamics. The moments of the
irreversibility pdf can be written explicitly and the pdf is seen to
be non-Gaussian. For nonequilibrium steady-states, the average
irreversibility grows from zero as the trajectory timespan increases,
with the initial growth rate being equal to the average entropy
production rate, which is related to the noise-amplification. In
nonequilibrium steady-states, there is
a finite time where the average irreversibility is maximal, and it
decays back to zero as time goes to infinity. 
For a 
system with a separation of timescales in the deterministic dynamics
and without asymptotically large noise in the fast modes,
only the slow modes contribute to the irreversibility statistics,
while modes faster than the trajectory timespan contribute higher order
corrections. There is, however, evidence that in at least some climate
subsystems the noise in fast modes is indeed large, and thus this
approach to dimensional reduction may not be applicable
\cite{PenlandSardeshmukh95,CompoSardeshmukhPenland2001}.

It is important to note that the irreversibility depends the
multivariate nature of both the deterministic dynamics and the random
noise. Nonequilibrium, the violation of detailed balance, and the
non-commutivity relation $\mA \mD - \mD \mA^T \ne 0$ are all
equivalent, and determining the irreversibility requires knowledge of
both parts of the dynamics. 

These results were obtained in the idealized context of a linear
dynamical system forced by additive Gaussian white noise. However, the
definition of the irreversibility and the constraint provided by the
fluctuation theorem are general. Thus, one can ask similar questions
about the distribution of irreversibility in more complex systems. It
remains to be seen which of the above results generalize. In
the context of climate, the fact that many climate subsystems can be
modeled by such simple stochastic dynamical systems indicates that the
properties obtained here are at least approximately valid for these
climate subsystems. Further, even for climate phenomena that are not
well-approximated by such simple models, the behavior of these models
provides a null-hypothesis for the phenomenon.

Fluctuations in climate subsystems are complex, multi-dimensional
phenomena with many characteristics. It is often convenient to reduce
this behavior down to a single scalar, referred to as an index. These
indices are usually based on subjectively chosen phenomenological
features of the fluctuations. For example,
one common index used in El-Ni\~no studies is the NINO3 index, defined
as the mean sea surface temperature anomaly from climatology over the
region 5$^\circ$N-5$^\circ$S, 90$^\circ$W-150$^\circ$W \cite{Trenberth1997}.
Different indices capture different aspects of a phenomenon,
and if the character of the phenomenon changes as climate changes,
then any particular index may lose its utility. 

The irreversibility provides an interesting alternative to traditional
indices. Like an index, irreversibility is a scalar defined in terms
of the time series of the system. However, unlike traditional indices,
the irreversibility reflects fundamental properties of the
nonequilibrium dynamics. The timescale of climate phenomena is often
understood in terms of the specific physical properties of the system
of interest. El-Ni\~no timescales, for example, involve the time for
waves to propagate across the tropical Pacific Ocean. Other timescales
such as the growth rates of perturbations based on singular vector
analysis, depend on the subjective coordinate system used. By
virtue of its coordinate independence, the time of maximum average
irreversibility provides an objective choice for a fluctuation
timescale. 

A major obstacle to the practical use of irreversibility is that its direct
measurement from Eq.\ (\ref{eq:rdef}) will require extremely long
timeseries. In order to compute $r$ directly from a timeseries, one must
quantify the probability of finding rare events, which means one must
have a time series long enough to contain those rare events. Even with
the long timeseries produced by numerical models, direct computation
of the irreversibility may be prohibitive for all but the very
simplest 
models. There are techniques that have been used to
accelerate nonequilibrium computations, but it remains to be seen if
these techniques can be applied to climate models. The most promising avenue
may be to use climate timeseries to estimate the parameters of the
linear stochastic model in Eq.\ (\ref{eq:sde}). This is the
technique used, for example, in constructing linear inverse models of
El-Ni\~no \cite{Penland93}. Then, the irreversibility could be
computed using the expressions obtained here. 

In many climate subsystems the phenomena modeled here by stationary
random noise have a strong seasonal component. A better stochastic model for
these phenomena would then be cyclostationary noise. Additional
complexities to be considered are red-noise processes and
multiplicative noise processes. All of these modifications will
probably impact the results obtained here.

As climate changes, the nonequilibrium steady-state
changes. Many climate subsystems have timescales shorter
than the timescale for climate change, and the change in the
steady-state can be considered to be adiabatic. By obtaining relations
between the parameters defining the steady-state and the fluctuations,
we have solved part of the question of how climate subsystem fluctuations will
be affected by climate change. The question of how the steady-state itself
evolves under climate change remains. However, it may be possible that
climate models do a better job of capturing the evolving steady state
than the fluctuations. If so, this work may lead to improved climate
change forecasts.

\begin{acknowledgments}
We wish to acknowledge Antonello Provenzale for useful discussions.
\end{acknowledgments}


\begin{thebibliography}{34}
\expandafter\ifx\csname natexlab\endcsname\relax\def\natexlab#1{#1}\fi
\expandafter\ifx\csname bibnamefont\endcsname\relax
  \def\bibnamefont#1{#1}\fi
\expandafter\ifx\csname bibfnamefont\endcsname\relax
  \def\bibfnamefont#1{#1}\fi
\expandafter\ifx\csname citenamefont\endcsname\relax
  \def\citenamefont#1{#1}\fi
\expandafter\ifx\csname url\endcsname\relax
  \def\url#1{\texttt{#1}}\fi
\expandafter\ifx\csname urlprefix\endcsname\relax\def\urlprefix{URL }\fi
\providecommand{\bibinfo}[2]{#2}
\providecommand{\eprint}[2][]{\url{#2}}

\bibitem[{\citenamefont{Evans and Searles}(2002)}]{EvansSearles2002}
\bibinfo{author}{\bibfnamefont{D.~J.} \bibnamefont{Evans}} \bibnamefont{and}
  \bibinfo{author}{\bibfnamefont{D.~J.} \bibnamefont{Searles}},
  \bibinfo{journal}{Adv. Phys.} \textbf{\bibinfo{volume}{51}},
  \bibinfo{pages}{1529} (\bibinfo{year}{2002}).

\bibitem[{\citenamefont{Searles and Evans}(1999)}]{SearlesEvans1999}
\bibinfo{author}{\bibfnamefont{D.~J.} \bibnamefont{Searles}} \bibnamefont{and}
  \bibinfo{author}{\bibfnamefont{D.~J.} \bibnamefont{Evans}},
  \bibinfo{journal}{Phys. Rev. E} \textbf{\bibinfo{volume}{60}},
  \bibinfo{pages}{159} (\bibinfo{year}{1999}).

\bibitem[{\citenamefont{Chernyak et~al.}(2006)\citenamefont{Chernyak, Chertkov,
  and Jarzynski}}]{Chernyak2006}
\bibinfo{author}{\bibfnamefont{V.~Y.} \bibnamefont{Chernyak}},
  \bibinfo{author}{\bibfnamefont{M.}~\bibnamefont{Chertkov}}, \bibnamefont{and}
  \bibinfo{author}{\bibfnamefont{C.}~\bibnamefont{Jarzynski}},
  \bibinfo{journal}{J. Stat. Mech.} \textbf{\bibinfo{volume}{P08001}}
  (\bibinfo{year}{2006}).

\bibitem[{\citenamefont{Penland and Magorian}(1993)}]{Penland93}
\bibinfo{author}{\bibfnamefont{C.}~\bibnamefont{Penland}} \bibnamefont{and}
  \bibinfo{author}{\bibfnamefont{T.}~\bibnamefont{Magorian}},
  \bibinfo{journal}{J. Climate} \textbf{\bibinfo{volume}{6}},
  \bibinfo{pages}{1067} (\bibinfo{year}{1993}).

\bibitem[{\citenamefont{Penland and Sardeshmukh}(1995)}]{PenlandSardeshmukh95}
\bibinfo{author}{\bibfnamefont{C.}~\bibnamefont{Penland}} \bibnamefont{and}
  \bibinfo{author}{\bibfnamefont{P.~D.} \bibnamefont{Sardeshmukh}},
  \bibinfo{journal}{J. Climate} \textbf{\bibinfo{volume}{8}},
  \bibinfo{pages}{1999} (\bibinfo{year}{1995}).

\bibitem[{\citenamefont{Moore and Kleeman}(1996)}]{MooreKleeman96}
\bibinfo{author}{\bibfnamefont{A.~M.} \bibnamefont{Moore}} \bibnamefont{and}
  \bibinfo{author}{\bibfnamefont{R.}~\bibnamefont{Kleeman}},
  \bibinfo{journal}{Q.J.R. Meteorol. Soc.} \textbf{\bibinfo{volume}{122}},
  \bibinfo{pages}{1405} (\bibinfo{year}{1996}).

\bibitem[{\citenamefont{Moore and Farrell}(1993)}]{MooreFarrell93}
\bibinfo{author}{\bibfnamefont{A.~M.} \bibnamefont{Moore}} \bibnamefont{and}
  \bibinfo{author}{\bibfnamefont{B.~F.} \bibnamefont{Farrell}},
  \bibinfo{journal}{J. Phys. Oceanogr.} \textbf{\bibinfo{volume}{23}},
  \bibinfo{pages}{1682} (\bibinfo{year}{1993}).

\bibitem[{\citenamefont{Farrell and Ioannou}(1993)}]{FarrellIoannou93}
\bibinfo{author}{\bibfnamefont{B.~F.} \bibnamefont{Farrell}} \bibnamefont{and}
  \bibinfo{author}{\bibfnamefont{P.~J.} \bibnamefont{Ioannou}},
  \bibinfo{journal}{J. Atmos. Sci.} \textbf{\bibinfo{volume}{50}},
  \bibinfo{pages}{4044} (\bibinfo{year}{1993}).

\bibitem[{\citenamefont{Farrell and Ioannou}(1994)}]{FarrellIoannou94}
\bibinfo{author}{\bibfnamefont{B.~F.} \bibnamefont{Farrell}} \bibnamefont{and}
  \bibinfo{author}{\bibfnamefont{P.~J.} \bibnamefont{Ioannou}},
  \bibinfo{journal}{J. Atmos. Sci.} \textbf{\bibinfo{volume}{51}},
  \bibinfo{pages}{2685} (\bibinfo{year}{1994}).

\bibitem[{\citenamefont{Newman et~al.}(1997)\citenamefont{Newman, Sardeshmukh,
  and Penland}}]{Newmanetal97}
\bibinfo{author}{\bibfnamefont{M.}~\bibnamefont{Newman}},
  \bibinfo{author}{\bibfnamefont{P.~D.} \bibnamefont{Sardeshmukh}},
  \bibnamefont{and} \bibinfo{author}{\bibfnamefont{C.}~\bibnamefont{Penland}},
  \bibinfo{journal}{J. Atmos. Sci.} \textbf{\bibinfo{volume}{54}},
  \bibinfo{pages}{435} (\bibinfo{year}{1997}).

\bibitem[{\citenamefont{Whitaker and
  Sardeshmukh}(1998)}]{WhitakerSardeshmukh98}
\bibinfo{author}{\bibfnamefont{J.~S.} \bibnamefont{Whitaker}} \bibnamefont{and}
  \bibinfo{author}{\bibfnamefont{P.~D.} \bibnamefont{Sardeshmukh}},
  \bibinfo{journal}{J. Atmos. Sci.} \textbf{\bibinfo{volume}{55}},
  \bibinfo{pages}{237} (\bibinfo{year}{1998}).

\bibitem[{\citenamefont{Weickmann et~al.}(2000)\citenamefont{Weickmann,
  Robinson, and Penland}}]{Weickmannetal00}
\bibinfo{author}{\bibfnamefont{K.~M.} \bibnamefont{Weickmann}},
  \bibinfo{author}{\bibfnamefont{W.~A.} \bibnamefont{Robinson}},
  \bibnamefont{and} \bibinfo{author}{\bibfnamefont{C.}~\bibnamefont{Penland}},
  \bibinfo{journal}{J. Geophys. Res.} \textbf{\bibinfo{volume}{105}},
  \bibinfo{pages}{15543} (\bibinfo{year}{2000}).

\bibitem[{\citenamefont{Penland and Matrosova}(2001)}]{PenlandMatrosova01}
\bibinfo{author}{\bibfnamefont{C.}~\bibnamefont{Penland}} \bibnamefont{and}
  \bibinfo{author}{\bibfnamefont{L.}~\bibnamefont{Matrosova}},
  \bibinfo{journal}{MWR} \textbf{\bibinfo{volume}{129}}, \bibinfo{pages}{1740}
  (\bibinfo{year}{2001}).

\bibitem[{\citenamefont{Farrell and
  Ioannou}(1996{\natexlab{a}})}]{FarrellgentheoryI}
\bibinfo{author}{\bibfnamefont{B.~F.} \bibnamefont{Farrell}} \bibnamefont{and}
  \bibinfo{author}{\bibfnamefont{P.~J.} \bibnamefont{Ioannou}},
  \bibinfo{journal}{J. Atmos. Sci.} \textbf{\bibinfo{volume}{53}},
  \bibinfo{pages}{2025} (\bibinfo{year}{1996}{\natexlab{a}}).

\bibitem[{\citenamefont{Farrell and
  Ioannou}(1996{\natexlab{b}})}]{FarrellgentheoryII}
\bibinfo{author}{\bibfnamefont{B.~F.} \bibnamefont{Farrell}} \bibnamefont{and}
  \bibinfo{author}{\bibfnamefont{P.~J.} \bibnamefont{Ioannou}},
  \bibinfo{journal}{J. Atmos. Sci.} \textbf{\bibinfo{volume}{53}},
  \bibinfo{pages}{2041} (\bibinfo{year}{1996}{\natexlab{b}}).

\bibitem[{\citenamefont{Ioannou}(1995)}]{Ioannou95}
\bibinfo{author}{\bibfnamefont{P.~J.} \bibnamefont{Ioannou}},
  \bibinfo{journal}{J. Atmos. Sci.} \textbf{\bibinfo{volume}{52}},
  \bibinfo{pages}{1155} (\bibinfo{year}{1995}).

\bibitem[{\citenamefont{Weiss}(2003)}]{Weiss2003}
\bibinfo{author}{\bibfnamefont{J.~B.} \bibnamefont{Weiss}},
  \bibinfo{journal}{Tellus} \textbf{\bibinfo{volume}{55A}},
  \bibinfo{pages}{208} (\bibinfo{year}{2003}).

\bibitem[{\citenamefont{Lorenz}(1965)}]{Lorenz65}
\bibinfo{author}{\bibfnamefont{E.~N.} \bibnamefont{Lorenz}},
  \bibinfo{journal}{Tellus} \textbf{\bibinfo{volume}{17}}, \bibinfo{pages}{321}
  (\bibinfo{year}{1965}).

\bibitem[{\citenamefont{van Oldenborgh et~al.}(2005)\citenamefont{van
  Oldenborgh, S.~Y.~Philip, and Collins}}]{vanOldenborgh2005}
\bibinfo{author}{\bibfnamefont{G.~J.} \bibnamefont{van Oldenborgh}},
  \bibinfo{author}{\bibfnamefont{S.}~\bibnamefont{S.~Y.~Philip}},
  \bibnamefont{and} \bibinfo{author}{\bibfnamefont{M.}~\bibnamefont{Collins}},
  \bibinfo{journal}{Ocean Science} \textbf{\bibinfo{volume}{1}},
  \bibinfo{pages}{81} (\bibinfo{year}{2005}).

\bibitem[{\citenamefont{Vallis}(2006)}]{Vallis2006}
\bibinfo{author}{\bibfnamefont{G.~K.} \bibnamefont{Vallis}},
  \emph{\bibinfo{title}{Atmospheric and Oceanic Fluid Dynamics}}
  (\bibinfo{publisher}{Cambridge University Press, Cambridge},
  \bibinfo{year}{2006}).

\bibitem[{\citenamefont{Penland}(1996)}]{Penland96}
\bibinfo{author}{\bibfnamefont{C.}~\bibnamefont{Penland}},
  \bibinfo{journal}{Physica D} \textbf{\bibinfo{volume}{98}},
  \bibinfo{pages}{534} (\bibinfo{year}{1996}).

\bibitem[{\citenamefont{C.E.Leith}(1996)}]{Leith1996}
\bibinfo{author}{\bibnamefont{C.E.Leith}}, \bibinfo{journal}{Physica D}
  \textbf{\bibinfo{volume}{98}}, \bibinfo{pages}{481} (\bibinfo{year}{1996}).

\bibitem[{\citenamefont{Majda et~al.}(1999)\citenamefont{Majda, Timofeyev, and
  Vanden-Eijnden}}]{MTV1999}
\bibinfo{author}{\bibfnamefont{A.~J.} \bibnamefont{Majda}},
  \bibinfo{author}{\bibfnamefont{I.}~\bibnamefont{Timofeyev}},
  \bibnamefont{and}
  \bibinfo{author}{\bibfnamefont{E.}~\bibnamefont{Vanden-Eijnden}},
  \bibinfo{journal}{Proc. Natl. Acad. Sci. USA} \textbf{\bibinfo{volume}{96}},
  \bibinfo{pages}{14687} (\bibinfo{year}{1999}).

\bibitem[{\citenamefont{Majda et~al.}(2001)\citenamefont{Majda, Timofeyev, and
  Vanden-Eijnden}}]{MTV2001}
\bibinfo{author}{\bibfnamefont{A.~J.} \bibnamefont{Majda}},
  \bibinfo{author}{\bibfnamefont{I.}~\bibnamefont{Timofeyev}},
  \bibnamefont{and}
  \bibinfo{author}{\bibfnamefont{E.}~\bibnamefont{Vanden-Eijnden}},
  \bibinfo{journal}{Commun. Pure App. Math.} \textbf{\bibinfo{volume}{54}},
  \bibinfo{pages}{891} (\bibinfo{year}{2001}).

\bibitem[{\citenamefont{Majda et~al.}(2006)\citenamefont{Majda, Timofeyev, and
  Vanden-Eijnden}}]{MTV2006}
\bibinfo{author}{\bibfnamefont{A.~J.} \bibnamefont{Majda}},
  \bibinfo{author}{\bibfnamefont{I.}~\bibnamefont{Timofeyev}},
  \bibnamefont{and}
  \bibinfo{author}{\bibfnamefont{E.}~\bibnamefont{Vanden-Eijnden}},
  \bibinfo{journal}{Nonlinearity} \textbf{\bibinfo{volume}{19}},
  \bibinfo{pages}{769} (\bibinfo{year}{2006}).

\bibitem[{\citenamefont{Seifert}(2005)}]{Seifert2005}
\bibinfo{author}{\bibfnamefont{U.}~\bibnamefont{Seifert}},
  \bibinfo{journal}{Phys. Rev. Lett.} \textbf{\bibinfo{volume}{95}},
  \bibinfo{pages}{040602} (\bibinfo{year}{2005}).

\bibitem[{\citenamefont{Brooks}(2005)}]{MatrixRefMan}
\bibinfo{author}{\bibfnamefont{M.}~\bibnamefont{Brooks}},
  \emph{\bibinfo{title}{The matrix reference manual}} (\bibinfo{year}{2005}),
  \urlprefix\url{http://www.ee.ic.ac.uk/hp/staff/dmb/matrix/intro.html}.

\bibitem[{\citenamefont{Abramowitz and Stegun}(1975)}]{AbramSteg}
\bibinfo{author}{\bibfnamefont{M.}~\bibnamefont{Abramowitz}} \bibnamefont{and}
  \bibinfo{author}{\bibfnamefont{I.~A.} \bibnamefont{Stegun}},
  \emph{\bibinfo{title}{Handbook of mathematical functions with formulas,
  graphs, and mathematical tables}} (\bibinfo{publisher}{Academic Press},
  \bibinfo{year}{1975}).

\bibitem[{\citenamefont{Carberry et~al.}(2004)\citenamefont{Carberry, Reid,
  Wang, Sevick, Searles, and Evans}}]{CarberryEtAl2004}
\bibinfo{author}{\bibfnamefont{D.~M.} \bibnamefont{Carberry}},
  \bibinfo{author}{\bibfnamefont{J.~C.} \bibnamefont{Reid}},
  \bibinfo{author}{\bibfnamefont{G.~M.} \bibnamefont{Wang}},
  \bibinfo{author}{\bibfnamefont{E.~M.} \bibnamefont{Sevick}},
  \bibinfo{author}{\bibfnamefont{D.~J.} \bibnamefont{Searles}},
  \bibnamefont{and} \bibinfo{author}{\bibfnamefont{D.~J.} \bibnamefont{Evans}},
  \bibinfo{journal}{Phys. Rev. Lett.} \textbf{\bibinfo{volume}{92}}
  (\bibinfo{year}{2004}).

\bibitem[{\citenamefont{Ritort}(2004)}]{Ritort2004}
\bibinfo{author}{\bibfnamefont{F.}~\bibnamefont{Ritort}}, \bibinfo{journal}{J.
  Stat. Mech.: Theor. Exp.} p. \bibinfo{pages}{P10016} (\bibinfo{year}{2004}).

\bibitem[{\citenamefont{Blickle et~al.}(2006)\citenamefont{Blickle, Speck,
  Helden, Seifert, and Bechinger}}]{BlickleEtAl2006}
\bibinfo{author}{\bibfnamefont{V.}~\bibnamefont{Blickle}},
  \bibinfo{author}{\bibfnamefont{T.}~\bibnamefont{Speck}},
  \bibinfo{author}{\bibfnamefont{L.}~\bibnamefont{Helden}},
  \bibinfo{author}{\bibfnamefont{U.}~\bibnamefont{Seifert}}, \bibnamefont{and}
  \bibinfo{author}{\bibfnamefont{C.}~\bibnamefont{Bechinger}},
  \bibinfo{journal}{Phy.s Rev. Lett} \textbf{\bibinfo{volume}{96}},
  \bibinfo{pages}{070603} (\bibinfo{year}{2006}).

\bibitem[{\citenamefont{Crooks and Jarzynski}(2007)}]{CrooksJarzynski2007}
\bibinfo{author}{\bibfnamefont{G.~E.} \bibnamefont{Crooks}} \bibnamefont{and}
  \bibinfo{author}{\bibfnamefont{C.}~\bibnamefont{Jarzynski}},
  \bibinfo{journal}{Phys. Rev. E} \textbf{\bibinfo{volume}{75}},
  \bibinfo{pages}{02116} (\bibinfo{year}{2007}).

\bibitem[{\citenamefont{Compo et~al.}(2001)\citenamefont{Compo, Sardeshmukh,
  and Penland}}]{CompoSardeshmukhPenland2001}
\bibinfo{author}{\bibfnamefont{G.~P.} \bibnamefont{Compo}},
  \bibinfo{author}{\bibfnamefont{P.~D.} \bibnamefont{Sardeshmukh}},
  \bibnamefont{and} \bibinfo{author}{\bibfnamefont{C.}~\bibnamefont{Penland}},
  \bibinfo{journal}{J. Climate} \textbf{\bibinfo{volume}{14}},
  \bibinfo{pages}{3356} (\bibinfo{year}{2001}).

\bibitem[{\citenamefont{Trenberth}(1997)}]{Trenberth1997}
\bibinfo{author}{\bibfnamefont{K.~E.} \bibnamefont{Trenberth}},
  \bibinfo{journal}{Bull. Am. Met. Soc.} \textbf{\bibinfo{volume}{78}},
  \bibinfo{pages}{2771} (\bibinfo{year}{1997}).

\end{thebibliography}
\end{document}